\documentclass[a4paper,11pt]{article}

\usepackage{fullpage}
\usepackage[T1]{fontenc}

\usepackage{amssymb,amsmath,cite}
\usepackage{hyperref,comment,xcolor}

\usepackage{amsthm}

\begin{document}
\vspace{5mm}
\vspace{0.5cm}

\def\thefootnote{\arabic{footnote}}
\setcounter{footnote}{0}

\allowdisplaybreaks 

\begin{titlepage}
\thispagestyle{empty}
\begin{flushright}
	\hfill{DFPD-2017/TH/04} 
\end{flushright}
				
\vspace{35pt}
				
\begin{center}
	{\Large{\bf From Linear to Non-linear SUSY  
	\\[0.3 cm] 
	and Back Again }}
									
	\vspace{50pt}
							
	{N.~Cribiori, G.~Dall'Agata and	 F.~Farakos}
							
	\vspace{25pt}
							
	{
		{\it  Dipartimento di Fisica e Astronomia ``Galileo Galilei''\\
			Universit\`a di Padova, Via Marzolo 8, 35131 Padova, Italy}
										
		\vspace{15pt}
										
		{\it   INFN, Sezione di Padova \\
		Via Marzolo 8, 35131 Padova, Italy}
		}
								
\vspace{40pt}
								
{ABSTRACT} 
\end{center}

We show that couplings of the goldstino field, appearing when global supersymmetry is broken, can always be described in superspace by means of a nilpotent chiral superfield $X$, satisfying $X^2=0$. This applies to both F-term and D-term supersymmetry breaking, even when the supersymmetry breaking model admits a linear description.

\vspace{10pt}
			
\bigskip
			
\end{titlepage}

\numberwithin{equation}{section}

\baselineskip 6 mm

\tableofcontents

\section{Introduction} 

Non-linear realizations can be a very useful tool to organize and constrain interactions in models where a global symmetry has been spontaneously broken \cite{Callan:1969sn}. 
The generator of the broken symmetry is associated to a goldstone mode, which is used to implement the non-linear realization by constructing appropriate interactions with the other fields and whose transformation law depends crucially on the energy scale at which the linear symmetry is restored. 
In general, when considering a physical system with a spontaneously broken symmetry, non-linear realizations emerge at energy scales below the scale set by the operator generating such symmetry breaking. 
In the case of supersymmetry this is the scale generated by non-vanishing F- or D-terms, which in turn sets the scale of the vacuum potential. 
As a consequence of supersymmetry breaking, one generically expects mass splittings in the spectrum, with some heavy states getting masses close to or even higher than the supersymmetry breaking scale. 
In such a scenario, in the low energy regime, one might integrate out such states and obtain an effective theory where supersymmetry must be realized non-linearly on states that are not organized in multiplets anymore.

Various approaches have been developed to implement the non-linear realization of supersymmetry, initially in the component fields formulation and subsequently in superspace \cite{Volkov:1972jx,Volkov:1973ix,Rocek:1978nb,IK1,Lind,Samuel:1982uh,Casalbuoni:1988xh}. 
Superspace methods, in particular, are attractive for the formulation of low energy theories because they make various properties of supersymmetry manifest even when the spectrum is not supersymmetric anymore. 
In this context, an approach which has gained particular attention is the description through constrained superfields \cite{Rocek:1978nb,Casalbuoni:1988xh,Brignole:1997pe,Komargodski:2009rz}. 
While superfields  provide linear representations of supersymmetry, imposing additional supersymmetric constraints on them forces some of their components to be composite states of the elementary fields. 
This in turn implies that supersymmetry enjoys a linear realization only on composite states, while it is non-linearly realized on the elementary fields. 

Generic models may contain full superfields as well as constrained ones, but one must always have in the spectrum a goldstino field, which has been often described by means of a nilpotent chiral superfield $X$ satisfying the $X^2=0$ constraint. 
This constraint allows for non-trivial solutions every time supersymmetry is broken by vacua where its highest auxiliary field component $F^X$ is non-vanishing. 
Additional constraints can then be imposed on the matter sector \cite{Brignole:1997pe,Komargodski:2009rz,Luo:2010zp,
DallAgata:2015pdd,Ferrara:2016een,DallAgata:2016syy}.

Non-linear realizations of supersymmetry can be relevant in any setup where supersymmetry is spontaneously broken. 
In this spirit, constrained superfields have been used for the effective description of inflationary models that are embedded in supergravity \cite{Antoniadis:2014oya,Ferrara:2014kva,Kallosh:2014via,DallAgata:2014qsj,Ferrara:2015tyn,Carrasco:2015iij,DallAgata:2015zxp,Kahn:2015mla,Dudas:2016eej,Kallosh:2016ndd,McDonough:2016der,Hasegawa:2017hgd}, but also to study more general brane supersymmetry breaking scenarios in string theory reductions to four dimensions.
While explicit string constructions based on tachyon-free non-BPS systems are known since some time \cite{Antoniadis:1999xk,Angelantonj:1999jh,Aldazabal:1999jr} and their non-linear supersymmetry has been studied in detail \cite{Dudas:2000nv,Pradisi:2001yv}, their connection with the constrained superfields are still not fully understood  \cite{Bergshoeff:2015jxa,Kallosh:2015nia,Garcia-Etxebarria:2015lif,Dasgupta:2016prs,Vercnocke:2016fbt,Kallosh:2016aep,Aalsma:2017ulu}. 
This renewal of interest for non-linear realizations also led to various developments on the properties of the supersymmetry breaking sector in local supersymmetry 
\cite{Dudas:2015eha,Bergshoeff:2015tra,Hasegawa:2015bza,Antoniadis:2015ala,Kallosh:2015tea,Bandos:2015xnf,Cribiori:2016qif,Bandos:2016xyu,Benakli:2017whb}. 

Focusing on the nilpotent constraint on the superfield $X$, it has been shown that this constraint arises as a consequence of the decoupling of the sgoldstino, the scalar superpartner of the goldstino, from the low energy theory \cite{Casalbuoni:1988xh}. 
Even if in various known examples this decoupling can be performed in a smooth way, the generality of this description can be questioned \cite{Dudas:2011kt}. 
The aim of our work is to address this issue, which is central in the study of constrained superfields. 

We consider both F-term and D-term supersymmetry breaking and we show that, in any case, the sector that breaks supersymmetry can be described in terms of constrained superfields. 
In particular, in the approach we propose, each component field of a given superfield can be promoted to a constrained superfield or, vice versa, any unconstrained superfield arising from linear representations of supersymmetry can be parametrized as a combination of constrained ones. 
Using this approach we show that the goldstino interactions can always be consistently described by a chiral superfield $X$, satisfying $X^2=0$. 
Since the decoupling of the heavy modes in the IR is not going to alter the parametrization we propose, our approach shows \emph{de facto} that the low energy description of any spontaneously broken supersymmetric theory  contains the nilpotent chiral superfield $X$. 

Our results apply to globally supersymmetric gauged chiral models and they can be summarized as follows. 
When supersymmetry is spontaneously broken, we can always describe the goldstino field interactions by means of a chiral superfield $X$ satisfying
\begin{equation}
\nonumber
X^2=0 \, . 
\end{equation} 
When needed, any other component field can be eliminated from a given superfield $Q$ by imposing \cite{DallAgata:2016syy} 
\begin{equation} 
\nonumber
X \overline X \, Q = 0 \, ,  
\end{equation} 
with appropriate Lagrange multipliers \cite{Ferrara:2016een}. 
This can be used to reconstruct any desired field spectrum in the low energy regime as well as to find the appropriate parametrization of the linear representations in terms of the non-linear ones, for supersymmetry breaking Lagrangians.

We will mainly proceed by examples, but it is going to be evident that the outlined procedure is general and applicable to any system with broken global supersymmetry, no matter the F-term, D-term or mixed origin.

\section{The goldstino and the Ferrara--Zumino supercurrent}

Before presenting a detailed discussion of various scenarios with a general construction of effective Lagrangians where the supersymmetry breaking is parametrized by the nilpotent chiral superfield $X$, we would like to comment on what one should have expected, based on previous general studies, like the one in \cite{Komargodski:2009rz}. 

For any globally supersymmetric theory with K\"ahler potential $K$ and superpotential $W$, there is a supercurrent multiplet ${\cal J}_{\alpha \dot \alpha}$ that satisfies the 
conservation equation\footnote{We use the conventions of \cite{Wess:1992cp}.}  
\begin{equation}
\overline D^{\dot \alpha}  {\cal J}_{\alpha \dot \alpha} = D_\alpha \Xi \,  , 
\end{equation} 
where $\Xi$ is a chiral superfield given by the combination
\begin{equation}
\Xi =4 W - \frac13 \overline D^2 K \, . 
\end{equation} 
On quite general grounds one can see that the low-energy supercurrent is expressed in terms of the goldstino when supersymmetry is broken and, in turn, the $\theta$-component of $\Xi$ should be identified with the goldstino. 
When the sgoldstino $x = {\Xi}|_{\theta=0}$ is not present in the low-energy theory, then the operator $x$ creates composite states of the goldstino, fixing 
\begin{equation}
\Xi^2 =0 \, . 
\end{equation} 

However, one can see some situations when this is not straightforward. 
For instance, in the simple model
\begin{equation}
K = \Phi \overline \Phi \, , \qquad W = f  \, \Phi \, ,
\end{equation}
the shift symmetry $\phi\to\phi + c$, where $c$ is a complex constant, protects the scalar $\phi$ in $\Phi$, which remains massless also in the IR. 
This implies that  $\Xi = \frac{8}{3} f \Phi $ still contains a massless scalar and therefore 
\begin{equation}
\Xi^2 = \frac{64}{9} f^2 \Phi^2 \ne 0 \, .  
\end{equation}
There are also interesting models using constrained superfields where the goldstino is not aligned with the fermion in the quadratic nilpotent superfield $X$. 
For example, in the model with two orthogonal constrained superfields $X$ and $Y$, satisfying
\begin{equation}
\label{XY}
X^2 = 0 = XY \, , 
\end{equation}
this happens whenever one adds a linear term in $Y$ in the superpotential. 
For the simplest choice
\begin{equation}
K = X \overline X + Y \overline Y  \, , \qquad W = f  \, X + g \, Y \, ,
\end{equation}
one sees that $\Xi = \frac83 (f X +  g Y)$, which satisfies
\begin{equation}
\Xi^2 = \frac{64}{9} g^2 Y^2 \ne 0 \, . 
\end{equation}
In fact, \eqref{XY} implies $Y^3=0$, but $Y^2$ is still non-vanishing.

Finally, when supersymmetry is broken by a pure Fayet--Iliopoulos term $\xi$, we have
\begin{equation} 
\Xi = -\frac{\xi}{3} \overline D^2 V \, ,
\end{equation}
which again is not nilpotent.

While one may have models where $\Xi^2\neq 0$ also in the IR, we will show that we can always parametrize supersymmetry breaking by means of a constrained chiral superfield $X$, whose square vanishes. 
This may not contain the full goldstino, but it will always contribute to the goldstino interactions and its auxiliary field will be selected by the dominant source of supersymmetry breaking. 

\section{Chiral multiplets and F-term breaking} 

The first examples of supersymmetry breaking scenarios we analyze are models of interacting chiral multiplets, with F-term breaking. 
These models are very simple and they have been already analized previously \cite{Casalbuoni:1988xh,Komargodski:2009rz}, but our presentation is going to emphasize the salient features of our argument for the general existence of the nilpotent superfield $X$ in the low-energy effective theory.

The simplest model one can think of is given by a single chiral superfield $\Phi$, with a supersymmetry breaking linear superpotential 
\begin{equation}
\label{LPhilinear}
\mathcal{L}= \int d^4 \theta \, \Phi \overline \Phi + f\left(\int d^2 \theta\, \Phi + c.c. \right) .
\end{equation}
It is straightforward to see that supersymmetry is broken because $\langle F^\Phi \rangle = -4f \neq 0$ and that the fermionic component of $\Phi$ is the goldstino.  
As mentioned before, the sgoldstino (the scalar superpartner) remains massless, but one can easily amend this fact by introducing an additional operator in the Lagrangian suppressed by a scale $\Lambda > \sqrt f$, whose only net effect is to generate a mass for $\phi = \Phi|_{\theta=0}$, $i.e.$
\begin{equation}
\label{mu1}
-\frac{1}{\Lambda^2} \int d^4\theta\, \Phi^2 \overline \Phi^2.
\end{equation}
We therefore expect the low-energy effective theory at energy scales well below $\frac{f}{\Lambda}$ to be described by the goldstino alone. 
This is naturally realized by splitting the degrees of freedom of $\Phi$ between two constrained superfields $X$ and $S$, satisfying
\begin{equation}
\label{XH1}
X^2=0 \ ,\qquad X {\overline D}_{\dot \alpha} \overline S =0.
\end{equation} 
The first constraint removes the scalar from $X$ and expresses it in terms of its fermion $G$ and the auxiliary field $F$, delivering   
\begin{equation}
\label{Xexpansion}
X = \frac{G^2}{2F} + \sqrt{2}\, \theta^\alpha G_\alpha + \theta^2 F.
\end{equation}
The second constraint removes the fermionic component $\psi^S$ and the auxiliary field $F^S$ of $S$ from the spectrum, giving 
\begin{equation}
\label{Sexpansion}
S = s + \sqrt 2 i  \, \theta \sigma^m  \left( \frac{\overline G }{\overline F}  \right) \partial_m s + \theta^2 \left( \frac{\overline G^2}{2 \overline F^2} \partial^2 s 
- \partial_n \left( \frac{\overline G}{\overline F}  \right) \overline \sigma^m \sigma^n \frac{\overline G}{\overline F}  \partial_m s   \right) \, . 
\end{equation}
Using these superfields we can therefore define 
\begin{equation}
\label{FXH}
\Phi= X + S \, ,
\end{equation}
so that it contains the same degrees of freedom of the original unconstrained superfield. 
In other words, we parametrized the unconstrained chiral superfield $\Phi$ in terms of two constrained chiral superfields: the nilpotent superfield $X$, given by \eqref{Xexpansion}, and the sgoldstino superfield $S$, given by \eqref{Sexpansion}.
The important fact is that this decomposition does not depend on the UV/IR details of the model under consideration and therefore it can always be performed.

Before elaborating further the analysis and in order to clarify better the previous step, let us draw an analogy between our approach and the BEH mechanism.  
Consider for simplicity an SO($n$) Higgs scalar field $\vec {\cal H}$. 
The symmetry can be spontaneously broken by imposing $\langle \vec {\cal H}\rangle = f \vec n \neq 0$, where $|\vec n|=1$, and then the field can be parametrized by
\begin{equation}
\vec{\cal H} = (f + \sigma(x))\left[\sin \frac{\pi}{f}\frac{\vec \pi}{\pi} + \cos\frac{\pi}{f} \vec{n}\right] \,.
\end{equation}
In this decomposition both the massless goldstone modes $\vec \pi$ and the massive field $\sigma(x)$ are present in the theory. 
Our approach is essentially the supersymmetric analogue of this decomposition as well as of the subsequent construction of the effective theory for the $\vec \pi$ fields, expanded in operators suppressed by the symmetry breaking scale $f$, after $\sigma$ has been integrated out.

The Lagrangian of the model \eqref{LPhilinear}, with the mass term \eqref{mu1}, can be expressed in terms of the parametrization \eqref{FXH} as
\begin{equation}
\label{LLXH}
\mathcal{L} = \int d^4 \theta \left(X\overline X + S \overline S - \frac{1}{\Lambda^2}(4 X \overline X S \overline S + S^2 \overline S^2)\right) 
+ f \left(\int d^2 \theta (X + S)+c.c.\right). 
\end{equation}
This shows explicitly that a non-linearly realized theory of supersymmetry, with very specific couplings, like the ones in \eqref{LLXH}, can behave like a linearly realized one. 
However we can also use this form of the Lagrangian to efficiently describe the effective theory. 
If we probe energies well below $\frac{f}{\Lambda}$ we expect the massive scalar in the sgoldstino superfield $S$ to decouple and hence we can obtain the effective theory by setting $S=0$. 
In the zero momentum limit, the Lagrangian \eqref{LLXH} in component form becomes
\begin{equation}
\label{zp1}
{\cal L} = - f^2 + |F + f|^2 - 4 \frac{|F|^2}{\Lambda^2} | s |^2  \, ,
\end{equation} 
which is clearly minimized for configurations where
\begin{equation}
\label{s==0}
 s = 0 \,,
\end{equation}
which, because of (\ref{Sexpansion}), implies $S=0$. 
This in fact is not only a good effective theory but also a consistent truncation of the original model (at least in the zero momentum limit) and leads to the expected Volkov--Akulov (VA) model
\begin{equation}
	\label{VA}
	{\cal L} = \int d^4 \theta\, X \overline X + f \left(\int d^2 \theta\, X + c.c.\right)\,.
\end{equation}
Had we studied a more complicated UV model for $\Phi$, the only change would have been in the expression for the decoupling of $s$, which would have been more complicated than \eqref{s==0}. The low energy Lagrangian, indeed, would still have been expressed solely in terms of the nilpotent $X$, albeit in a more complicated form than \eqref{VA}. 

An alternative way of deriving the same effective model is by assuming that the operator \eqref{mu1} dominates the equations of motion, so that their superspace formulation becomes
\begin{equation}\label{eomphi}
\Phi \overline D^2 \overline \Phi^2 =0.
\end{equation}
Decomposing $\Phi$ as in \eqref{FXH} and acting with $X\overline X D^2$ gives
\begin{equation}
X\overline X \left( \overline S |D^2 X|^2 + 8 S \partial^2 \overline S^2 + 16 |S|^2 \partial^2 \overline X \right) = 0 \,.
\end{equation}
We can then use this equation to find an expression for $X \overline X S$ in terms of operators including derivatives on $S$ and on $X$
\begin{equation}
	X \overline X \overline S = - 8 \, X \overline X S \left( \partial^2 \overline S^2 + 2 \overline S \partial^2 \overline X\right)/|D^2 X|^2 
\end{equation}
and using iteratively the resulting expression on the right hand side of the relation we finally obtain
\begin{equation}
X\overline X S=0
\end{equation} 
and therefore
\begin{equation}
 S=0.
\end{equation} 

Supersymmetry breaking in globally supersymmetric models is often related to $R$-symmetry breaking, which implies the existence of an $R$-axion in the effective theory.
A very simple model with such feature can be constructed by improving the Lagrangian \eqref{LPhilinear} with quite generic corrections, like
\begin{equation}
\label{mulambda}
+ \frac{\mu}{4\Lambda^2} \int d^4\theta\, \Phi^2 \overline \Phi^2 
- \frac{\lambda}{9\Lambda^4} \int d^4\theta\, \Phi^3 \overline \Phi^3 , 
\end{equation}
where $\mu$ and $\lambda$ are positive real constants.
The model clearly has an $R$-symmetry and the superfield $\Phi$ has $R$-charge 2. 
The scalar potential 
\begin{equation}
V = \frac{f^2}{1 + \frac{\mu}{\Lambda^2} \phi \overline \phi - \frac{\lambda}{\Lambda^4} \phi^2 \overline \phi^2 } \, 
\end{equation}
admits a stable vacuum at 
\begin{equation}
\langle \phi \overline \phi \rangle  = \frac{\mu}{2 \lambda} \Lambda^2 \equiv v^2 \, , 
\end{equation}
with 
\begin{equation}
	\langle V\rangle = \frac{f^2}{1+ \lambda \frac{v^4}{\Lambda^4}}.
\end{equation}
The spectrum at this vacuum consists of a massless real scalar, which is the $R$-axion, one real scalar, with mass $m^2 = 128 \, \frac{f^2}{\Lambda^2} \frac{\lambda^3 \mu}{\left(4 \lambda + \mu^2 \right)^3}$ and the goldstino.
An effective way to describe the low energy theory of this model follows from the parameterization 
\begin{equation}
\Phi = X + \left(v + {\cal A} \right) {\cal R}^2 \, , 
\end{equation}
where $X$ is the standard goldstino superfield and ${\cal A}$ and ${\cal R}$ are chiral constrained superfields satisfying
\begin{equation}
X {\cal A} = X \overline{\cal A}  \ , \qquad X \left( {\cal R} \overline{\cal R} -1 \right) = 0 \, . 
\end{equation} 
Notice that the ${\cal R}$ superfield satisfies also $X \overline D_{\dot \alpha} \overline {\cal R} = 0$, under the assumption that $\langle {\cal R} \rangle \ne 0$, which holds here. 
The chiral superfield ${\cal A}$ has vanishing $R$-charge, while the chiral superfield ${\cal R}$ has $R$-charge 1, and it is easily related to another standard constrained superfield $B$, satisfying $XB = X \overline B$, by ${\cal R} = e^{i B}$ \cite{Komargodski:2009rz,Dine:2009sw}.

For the lowest component fields of ${\cal A}$ and ${\cal R}$ we have 
\begin{equation}
{\cal A} | = h + {\cal O}(G,\overline G)  \ , \quad {\cal R}| = \text{e}^{ia} + {\cal O}(G,\overline G)  \, , 
\end{equation}
where $h$ is a real scalar, which becomes massive, and $a$ is the $R$-axion. 
In terms of this parametrization, the zero momentum limit of the Lagrangian becomes 
\begin{equation}
\begin{aligned}
\label{Lha}
\mathcal{L} = - \frac{f^2}{1+ \lambda \frac{v^4}{\Lambda^4}} +  \left(1+ \lambda \frac{v^4}{\Lambda^4}\right)\left|\overline F + \frac{f}{1+ \lambda \frac{v^4}{\Lambda^4}}\right|^2 
- \lambda \frac{|F|^2}{\Lambda^4} 
\left[ h^2 \left( 2 v + h \right)^2\right] \, . 
\end{aligned}
\end{equation}
This is clearly minimized when $h=0$, which, from supersymmetry, implies that the complete superfield ${\cal A}$ vanishes in the low energy, namely ${\cal A}=0$. 
The low energy effective theory is then described by the effective action
\begin{equation}
\begin{aligned}
{\cal L} = & \int d^4 \theta \left( |X|^2 + f_a^2 |{\cal R}|^2 \right) 
 + \left( \int d^2 \theta \left(\hat{f}  X + \tilde{f}  {\cal R}^2 \right) + c.c. \right) ,
\end{aligned}
\end{equation}
where we used the fact that $ X \overline{\cal R}^2$ is chiral and 
that $({\cal R} \overline{\cal R})^n - n^2\, {\cal R} \overline{\cal R}$ is the real part of a chiral function.
This action coincides with the one presented in \cite{Komargodski:2009rz,Dine:2009sw}, upon identification of the parameters in our original model as follows
\begin{equation}
	\hat{f} = \frac{f}{\sqrt{\alpha}}, \qquad f_a^2 = 4 v^2 \alpha, \qquad \tilde{f} = v f\,, \qquad \alpha = 1+ \lambda \left(\frac{v}{\Lambda}\right)^4.
\end{equation}
As expected from the general analysis in \cite{Dine:2009sw}, the expectation value of the superpotential is at the threshold of the bound
\begin{equation}
	|\langle W \rangle| \leq \frac12 f_a F,
\end{equation}
where, in our model, $|\langle W \rangle| = \tilde f$ and $F = \hat f$.

We showed that, in the case of F-term breaking, the constrained superfield $X$ satisfying $X^2=0$ appears in the low energy, either alone or with the constrained superfield $S$, which contains the sgoldstino. 
Up to this point however our discussion was devoted only to the supersymmetry breaking sector. 
We show now how to treat possible matter couplings. 
In general in the low energy regime there are going to be both complete and incomplete supermultiplets. 
A complete supermultiplet does not need to be parametrized using non-linear realizations of supersymmetry. 
In particular if only complete matter supermultiplets are present, solely the goldstino superfield is going to realize supersymmetry in a non-linear way \cite{Antoniadis:2010hs}. 
On the other hand if a given multiplet is incomplete, one can use the generic constraint 
\begin{equation}
\label{XXQ}
X\overline X Q = 0 \, , 
\end{equation}
proposed in \cite{DallAgata:2016syy}, to remove any unwanted component field from the spectrum. 
In any case, with the procedure we introduced above, one can parametrize both complete and not complete supermultiplets in terms of constrained superfields. 
In the remaining part of this section we discuss a simple example in this direction. 

Consider a model with two chiral superfields $\Phi$ and $\Sigma$. 
In general they can both contribute to the supersymmetry breaking, but we assume for simplicity that only $\Phi$ breaks supersymmetry. 
Since the scalar components of both the superfields $\Phi$ and $\Sigma$ are going to get a mass in the model we are going to consider, in terms of constrained superfields we can parametrize $\Phi$ and $\Sigma$ as 
\begin{equation}
\label{XHYH}
\begin{aligned}
\Phi & = X + S \, , 
\\
\Sigma & = Y + H \, , 
\end{aligned}
\end{equation}
where $X$ and $S$ are the same constrained superfields introduced before, while $Y$ and $H$ are chiral superfields satisfying \cite{Brignole:1997pe,Komargodski:2009rz} 
\begin{equation}
\label{XY=0}
X \, Y = 0 \, ,  \qquad  X {\overline D}_{\dot \alpha} \overline H =0 \, . 
\end{equation}
The component expansion for $Y$ is 
\begin{equation}
Y = \frac{G \psi}{\tilde F} - \frac{G^2}{2 F^2} \tilde F  + \sqrt 2 \, \theta \psi + \theta^2  \tilde F\, . 
\end{equation} 
The meaning of the decomposition \eqref{XHYH} is the following. 
The superfield $\Phi$ is decomposed in the goldstino superfield $X$ and in the sgoldstino superfield $S$ as in the previous example. 
The superfield $\Sigma$ instead is decomposed into the superfield $Y$, which contains the physical fermion $\psi$ and the scalar auxiliary component field $\tilde F$, and into the superfield $H$, which contains the scalar component field $h$. 
A simple model with spontaneously broken supersymmetry is 
\begin{equation}
\label{Model2}
{\cal L} =  \int d^4 \theta \left( |\Phi|^2 + |\Sigma|^2 
- \frac{|\Phi|^4}{\Lambda^2}
- \frac{|\Phi|^2 |\Sigma|^2}{\Lambda^2} \right) 
+ f \left(\int d^2 \theta \Phi +c.c.\right) \, .
\end{equation}
As in the previous example, after the replacement \eqref{XHYH} this Lagrangian can be written only in terms of constrained superfields
\begin{equation}
\label{LXSYH}
\begin{aligned}
{\cal L} = &  \int d^4 \theta \left( |X|^2 + |S|^2 + |Y + H|^2 
- 4 \frac{|X|^2 |S|^2}{\Lambda^2} -  \frac{|S|^4 }{\Lambda^2}
-\frac{|X+S|^2}{\Lambda^2} |Y + H|^2 \right)  \\[3mm] 
& + f \left(\int d^2 \theta (X +S) +c.c.\right) \, .  
\end{aligned}
\end{equation} 
In particular the pure $X$ sector has again the form \eqref{VA} and this fact is not going to change in the case in which also $Y$ contributes to the supersymmetry breaking. 

We can once more analyze the low-energy effective limit by first looking at the zero-momentum equations.
In component form, this gives the Lagrangian
\begin{equation}
\label{zp22}
{\cal L} = - f^2 + |F + f|^2 + |\tilde F|^2 - 4 \frac{|F|^2}{\Lambda^2} | s |^2  - \frac{|F h + \tilde F s |^2}{\Lambda^2}. 
\end{equation}  
It is once again clear that in the low-energy limit we will have that 
\begin{equation}
\label{hs=0}
h = 0 \qquad {\rm and} \qquad  s  = 0 \,,
\end{equation} 
which eventually imply $S=H=0$. 

As in the previous example, one could also obtain the same result by taking the formal limit in which the terms suppressed by $\Lambda$ are taken to be large.
The superspace equation of motion of $\Phi$ will then be the same as (\ref{eomphi}), which is solved by $S = 0$ once we decompose $\Phi = X+S$. 
The superspace equation of motion for $\Sigma$ in this limit is 
\begin{equation}
\Phi \, \overline D^2 \left( \overline \Sigma \, \overline \Phi \right) = 0 \, 
\end{equation} 
which, once the parametrization \eqref{XHYH} is used, reduces to
\begin{equation}
\label{XHX} 
X \, \overline D^2 \left( \overline H \, \overline X \right) = 0 \, , 
\end{equation}
where $S=0$ and $XY=0$ have been used. 
Due to the properties of $H$, equation \eqref{XHX} gives directly $X \overline H=0$, which implies 
\begin{equation}
H=0 \, . 
\end{equation}  

We understand that, with the superfield parametrization we propose, the decoupling of the massive states in the low energy regime can be obtained in a very efficient way. 
The equations of motion at zero momentum, in particular, have a straightforward solution.
Changing the parametrization, it is not guaranteed that the decoupling can still be performed and in general, even in the cases in which it can, the calculations are going to be more involved.

\section{A parametrization for the $X$ superfield}

Before proceeding with the analysis of the models with D-term breaking or with mixed sources of supersymmetry breaking, we are going to discuss a particular parametrization for the nilpotent superfield $X$ that is going to allow us to simplify the derivation of some of the results presented in the next section. 
Since this is a rather technical intermezzo, the reader interested in the physics of the supersymmetry breaking mechanisms and in the discussion of the resulting low energy effective theories can skip this section at first.

The nilpotent superfield $X$ contains as degrees of freedom a fermion field and a complex scalar auxiliary field.
We can split these degrees of freedom by introducing the chiral superfield
\begin{equation}\label{definitionZ}
Z = \overline D^2 \left( \frac{X\overline X}{D^2 X\,\overline D^2
\overline X} \right) ,
\end{equation}
which satisfies 
\begin{equation}
\label{constrZ}
Z^2 =0 , \qquad  Z \, \overline D^2 \overline Z = Z 
\end{equation}
and was first introduced to discuss the couplings of the goldstino in \cite{Rocek:1978nb}.
The constraints (\ref{constrZ}) imply that $Z$ contains only a fermion field, which is related to the fermion in $X$ by a field redefinition such that the original goldstino always appears in the fixed combination $G/F$.

In order to express $X$ in terms of $Z$, we can see that from (\ref{definitionZ}) follows straightforwardly the following relation between $X$ and $Z$ \cite{Liu:2010sk,Cribiori:2016hdz} 
\begin{equation}
X = Z \frac{D^2 X}{D^2 Z} \,.
\end{equation}
This tells us that $X$ is proportional to $Z$ times an antichiral superfield $D^2 X/D^2 Z$.
However, we can also prove that this is equivalent to the decomposition
\begin{equation}
\label{XZA}
X = \frac{Z}{2} \left({\cal A}_1  + i {\cal A}_2  \right) \, , 
\end{equation}
where ${\cal A}_i$ are chiral superfields, built from the chiral projections of the real and imaginary parts of $D^2 X/D^2Z$, namely
\begin{equation}
{\cal A}_1 =  \overline D^2 \left( \frac{\overline Z}{\overline D^2
\overline Z}
\left[  \frac{D^2 X}{D^2 Z} +  \frac{\overline D^2 \overline
X}{\overline D^2 \overline Z} \right]
\right)  
\end{equation}
and
\begin{equation}
{\cal A}_2 = - i \,  \overline D^2 \left( \frac{\overline Z}{\overline
D^2 \overline Z}
\left[  \frac{D^2 X}{D^2 Z} -  \frac{\overline D^2 \overline
X}{\overline D^2 \overline Z} \right]
\right)  \, .
\end{equation}
Indeed, the superfields ${\cal A}_1$ and ${\cal A}_2$ are manifestly chiral and they satisfy the equivalent constraints
\begin{equation}
\label{constrXAZA}
X ({\cal A}_i - \overline{\cal A}_i )=0\,,  \qquad Z ({\cal A}_i -
\overline{\cal A}_i)=0\,.
\end{equation}
In particular the only independent component in the chiral superfield ${\cal A}_i$ is the real scalar, which resides in its lowest component, namely
\begin{equation}
{\cal A}_i |_{\theta = \bar \theta = 0} = a_i + {\cal O}(G,\overline G) \,  .
\end{equation}
Up to now we isolated the independent degrees of freedom in $X$, namely one fermion and two real scalars, and we promoted them to constrained superfields $Z$, $A_i$. 
These constrained superfields can be treated as independent. 
This means that the nilpotent goldstino superfield $X$ can be decomposed into three constrained chiral superfields: one pure goldstino superfield $Z$, containing only one fermion, and two auxiliary superfields $\mathcal{A}_{1,2}$, each one containing one real (auxiliary) scalar.

Using now this parametrization \eqref{XZA} in the VA Lagrangian (\ref{VA}), we get
\begin{equation}
\label{LZA}
\mathcal{L} = \frac14 \int d^4 \theta \, Z \overline Z \left( |{\cal
A}_1|^2 + |{\cal A}_2|^2 \right)
+ \frac{f}{2} \left(\int d^2 \theta Z ({\cal A}_1 + i {\cal A}_2)
+c.c.\right) \, ,
\end{equation}
where for simplicity we took $f$ to be real.
In this case, the properties of the superfields $Z$ and ${\cal A}_2$ also imply the interesting fact that their combination drops from the superpotential 
\begin{equation}
\int d^2 \theta \left( i f Z  {\cal A}_2 \right)  + c.c.= -4 i f \,
\int d^4 \theta \, Z \overline Z ({\cal A}_2  - \overline{\cal A}_2  )
=  0 .
\end{equation}
The Lagrangian \eqref{LZA} simplifies then to
\begin{equation}
\label{LAA12}
\mathcal{L} = \frac14 \int d^4 \theta \, Z \overline Z \left( |{\cal
A}_1|^2 + |{\cal A}_2|^2 \right)
+ \frac{f}{2} \left(\int d^2 \theta Z {\cal A}_1 +c.c.\right) \,
\end{equation}
and the variation with respect to ${\cal A}_2$ gives\footnote{We refer the reader to appendix \ref{appA2} for a detailed proof.}
\begin{equation}
\label{A2=0}
{\cal A}_2 = 0 \, .
\end{equation}
This shows that ${\cal A}_2$ is a trivial auxiliary field, whose net effect on the calculation of the final Lagrangian is null.
It is then clear that the VA lagrangian is equivalent to
\begin{equation}
\label{onlyA1A1}
\mathcal{L} = \frac14 \int d^4 \theta \, |Z|^2 |{\cal A}_1|^2
+ \frac{f}{2} \left(\int d^2 \theta Z {\cal A}_1 +c.c.\right) \,
\end{equation}
and using the inverse relation for ${\cal A}_1$ in terms of $X$
\begin{equation}
Z {\cal A}_1 = X \left( 1 + \overline D^2 \left( \frac{\overline
X}{D^2 X } \right) \right) \, ,
\end{equation}
we can finally express this model in terms of the original superfield $X$
\begin{equation}
\label{onlyA1}
\mathcal{L}=\frac14 \int d^4 \theta \, X \overline X \left(2 + \frac{D^2 X}{\overline D^2 \overline X} + \frac{\overline D^2 \overline X}{D^2 X} \right) + \left(f \int d^2 \theta\, X+c.c.\right) .
\end{equation}
This Lagrangian contains two new terms with respect to the standard VA lagrangian which have the form of higher derivatives.
However, as we saw, the two lagrangians are effectively equivalent on shell and the new terms are present in order to cancel the degree of freedom encoded into the imaginary part of the auxiliary field $F$ of the superfield $X$.
Indeed the solution \eqref{A2=0} is telling us that such degree of freedom, which is associated precisely to $\mathcal{A}_2$, is not going to appear in the Lagrangian  \eqref{onlyA1}, because this Lagrangian is equivalent to \eqref{onlyA1A1}.
It is important to stress that this result does not imply that the imaginary part of $F$ is set to zero by the equations of motion: it is instead replaced by a composite expression built out of goldstini. 
Following the reverse reasoning, when the parameter $f$ is real the higher derivative terms in \eqref{onlyA1} can be eliminated by restoring the ${\cal A}_2$ part in the Lagrangian. 
This step can always be perfomed (at least at the classical level) because, as already noticed, the models \eqref{LAA12} and \eqref{onlyA1A1} are equivalent due to \eqref{A2=0}.
It is straightforward to show that 
\begin{equation}
Z {\cal A}_2 = -i\,X \left( 1 -\overline D^2 \left( \frac{\overline
X}{D^2 X } \right) \right) \, ,
\end{equation}
which implies
\begin{equation}
\label{ZZA2}
\frac14 \int d^4 \theta\, Z \overline Z |\mathcal{A}_2|^2 = \frac14 \int d^4 \theta \, X \overline X \left(2 -\frac{D^2 X}{\overline D^2 \overline X} - \frac{\overline D^2 \overline X}{D^2 X}
\right) \, ,
\end{equation}
which, added to \eqref{onlyA1}, completes the VA Lagrangian.
To recapitulate, we showed that, if the parameter $f$ is real, the VA model (\ref{VA})  and \eqref{onlyA1} are equivalent.

Before ending this section a comment is in order.
The previous discussion concerning the imaginary part of the auxiliary field $F$ involved only the pure goldstino sector, without taking into account the matter sector.
In particular one should ask if the solution \eqref{A2=0} holds also in more general cases, where matter superfields are present and some of their components might be removed.
To answer this question notice first that the generic way to eliminate matter component fields is described by constraints of the form given in \eqref{XXQ}.
Since the constraint \eqref{XXQ} is equivalent to
\begin{equation}
Z\overline Z Q =0 \, ,
\end{equation}
the eliminated components are not going to depend neither on the components of ${\cal A}_1$ nor on the ones of ${\cal A}_2$, but only on the components of $Z$.
Therefore the constraints on the matter superfields are independent of $a_2$ and the result presented here holds also in that case.

\section{Vector multiplets}

We now move on to the discussion of supersymmetry breaking scenarios involving vector multiplets.
First we consider the case of a pure D-term breaking and then we analyze more general situations in which a mixing between D-term and F-term breaking occurs. 
In particular we are going to show that, even for D-term supersymmetry breaking, the low energy theory can be parametrized in terms of a nilpotent chiral superfield $X$. 

\subsection{Pure D-term breaking}

Given a vector superfield $V$, a simple model realizing pure D-term supersymmetry breaking is
\begin{equation}
\label{LWV}
\mathcal{L} = \frac14 \left(\int d^2 \theta \, W^\alpha W_\alpha + c.c. \right) + \xi \int d^4\theta \,V,
\end{equation}
where $W_\alpha = -\frac14 \overline D^2 D_\alpha V$ is the vector field strength chiral superfield and $\xi$ is a Fayet--Iliopoulos parameter. 
Supersymmetry is broken whenever $\xi\neq 0$. 

In the spirit of the previous discussion, we want to parametrize this theory of spontaneously broken but linearly realized supersymmetry in terms of constrained superfields. 
To this purpose, we consider one chiral superfield $X$ and one real superfield $\tilde{V}$ such that
\begin{equation}
\label{constrXW}
X^2=0,\qquad X \tilde{W}_\alpha =0, \qquad X\overline X D^\alpha \tilde{W}_\alpha=0,
\end{equation}
where $\tilde{W}_\alpha=-\frac14 \overline D^2 D_\alpha \tilde V$.
The first constraint removes the scalar component from $X$, as in the previous discussion, while the second and the third constraints remove the fermion and the auxiliary field $\tilde {\rm D}$ from $\tilde{V}$. 
The only independent component field in $\tilde V$ is therefore a real vector field. 
The superspace expansion of $\tilde W_\alpha$ in particular is 
\begin{equation}
\label{Wtilde}
\tilde W_\alpha = - i \tilde \lambda_\alpha  + \tilde L_{\alpha}^{\beta} \theta_\beta  + \sigma^m_{\alpha \dot \beta} \, \partial_m \overline{\tilde \lambda}^{\dot \beta} \theta^2 \, ,  
\end{equation}
where 
\begin{equation}
\tilde L_{\alpha}^{\beta} =  \delta_{\alpha}^{\beta}\;  \tilde{\text{D}} - \frac{i}{2}\, (\sigma^m \overline \sigma^n)_{\alpha}^{\ \beta} F_{mn} \, 
\end{equation}
and $F_{mn}$ is the vector field-strength. 
Due to the constraints the gaugino $\tilde\lambda_\alpha$ and the auxiliary field $\tilde {\rm D}$ are expressed as composite combinations of the other degrees of freedom. 
Expanding in terms of powers of the goldstino field $G_{\alpha}$ one gets that
\begin{equation}
\begin{aligned}
\tilde \lambda_\alpha =& \frac{1}{2 \sqrt 2} (\sigma^m \overline \sigma^n)_{\alpha}^{\ \beta} \frac{G_\beta}{F} F_{mn} 
+ \cdots ,\\[3mm]
\tilde{\text{D}} =&  \frac12 \left[ \partial_c \left( \frac{G}{\sqrt 2 F} \right) \sigma^a \overline \sigma^b \sigma^c 
\left( \frac{\overline G}{\sqrt 2 \overline F} \right) \right] F_{ab} 
-  \frac12 \left[  \left( \frac{G}{\sqrt 2 F} \right) \sigma^c \overline \sigma^a \sigma^b 
\partial_c \left( \frac{\overline G}{\sqrt 2 \overline F} \right) \right] F_{ab} 
\\[2mm]
& -  \frac12 \left[ \left( \frac{G}{\sqrt 2 F} \right) \sigma^c \overline \sigma^a \sigma^b 
\left( \frac{\overline G}{\sqrt 2 \overline F} \right) \right] \partial_c F_{ab}  
+ \cdots  ,
\end{aligned}
\end{equation}
where dots stand for higher order goldstino terms. 

Using these constrained superfields we can introduce the parametrization
\begin{equation}
\label{VX}
V= \tilde V + \sqrt 2 \frac{X\overline X}{D^2 X}+ \sqrt 2 \frac{X\overline X}{\overline D^2 \overline X} \, , 
\end{equation}
which indeed contains the complete amount of degrees of freedom of an unconstrained vector superfield, namely one real vector, one fermion and one real scalar. 
This parametrization is the analogous of \eqref{FXH} for the case of pure D-term breaking. 
It is important to note that, if we also parametrize $X$ according to \eqref{XZA}, the contribution of $\mathcal{A}_2$ disappears from \eqref{VX}, because
\begin{equation}
\frac{X\overline X}{D^2 X} + \frac{X\overline X}{\overline D^2 \overline X} = 4 Z \overline Z \mathcal{A}_1.
\end{equation}
This was not obvious a priori but it is complementary to the result obtained in the previous section. 
In fact when parametrizing (part of) a vector superfield in terms of constrained chiral superfields, one clear obstacle arises because the auxiliary field of a vector superfield is real, while the auxiliary field of a chiral superfield is complex. 
The problem related to this mismatching of auxiliary degrees of freedom is immediately solved for all those systems in which the imaginary part of the auxiliary field $F$ of the superfield $X$ does not get an independent vacuum expectation value. 
As shown in the previous section, this occurs whenever the parameter $f$ giving the vacuum expectation value of $F$ is real and in general this condition is not too restrictive.

At this point we can insert the parametrization \eqref{VX} inside the Lagrangian \eqref{LWV} and, using the constraints \eqref{constrXW}, rewrite it as
\begin{equation}
\begin{aligned}
\label{LLXV2}
\mathcal{L}  &= \frac14 \left(\int d^2\theta\, \tilde W^\alpha \tilde W_\alpha +c.c.\right)+ \xi \int d^4 \theta\, \tilde V \, \\[3mm]
&+ \frac14 \int d^4 \theta \, X \overline X \left( 2 + \frac{D^2 X}{\overline D^2 \overline X} + \frac{\overline D^2 \overline X}{D^2 X} 
\right)  - \frac{\xi\sqrt 2}{4 }\left(\int d^2 \theta\, X+c.c.\right) . 
\end{aligned}
\end{equation}
As explained in the previous section, one can harmlessly introduce a new term proportional to (\ref{ZZA2}), which vanishes on-shell, so that the Lagrangian takes the more pleasant form
\begin{equation}
\label{LLXV}
\mathcal{L}  = \frac14 \left(\int d^2\theta\, \tilde W^\alpha \tilde W_\alpha +c.c.\right)+ \xi \int d^4 \theta\, \tilde V +  \int d^4 \theta \, X \overline X- \frac{\xi\sqrt 2}{4 }\left(\int d^2 \theta\, X+c.c.\right) . 
\end{equation}
The Lagrangian \eqref{LLXV} manifestly describes a theory of non-linearly realized supersymmetry, where the goldstino in $X$ interacts with the real vector field in $\tilde V$. 
This theory is completely equivalent to the original model \eqref{LWV}, where only the vector superfield $V$ is present and supersymmetry is broken by the Fayet--Iliopoulos term. 
In other words we showed that, for pure D-term supersymmetry breaking, the theory can be parametrized in terms of a nilpotent chiral superfield $X$ accomodating the goldstino.

This analysis can be easily generalized in order to include matter couplings and possibly an abelian gauge symmetry
\begin{equation}
\label{GEN1}
\begin{aligned}
\mathcal{L} = &  
\frac{1}{4} \left(\int d^2\theta \, {\cal F}(\Phi^I) \,  W^\alpha W_\alpha +c.c.\right) 
+ \xi \int d^4 \theta\, V 
+ \int d^4 \theta \sum_I \, \overline \Phi^I  \text{e}^{q_I V} \Phi^I 
\\[2mm]
& 
- \int d^4 \theta \sum_{IJ} \, \mu_{IJ} \left( \overline \Phi^I  \text{e}^{q_I V} \Phi^I \right) \left( \overline \Phi^J  \text{e}^{q_J V} \Phi^J \right)
+ \left(\int d^2 \theta\, {\cal W}(\Phi^I) + c.c. \right) \, .  
\end{aligned}
\end{equation} 
The detail of the parameterization of the various fields depends on the supersymmetry breaking pattern and especially on the low-energy spectrum.  
However we note that when scalar fields survive and we split some of the chiral fields as
\begin{equation}
	\Phi^I = H^I + Y^I,
\end{equation}
we will get an additional contribution to the superpotential of the form\footnote{Here we use the Wess--Zumino gauge $X \tilde V=0$ proposed in \cite{Komargodski:2009rz}. }
\begin{equation}
	 -\frac{\sqrt 2}{4} X \left[ \xi  	+ \sum_I q_I  |H^I|^2	+  \sum_{IJ} \mu_{IJ} (q_I + q_J)  |H^I|^2 |H^J|^2 \right] ,
\end{equation}
because $X \overline{D}_{\dot \alpha} \overline{H}^I = 0$. 
We should also note that for a generic choice of gauge kinetic function, the normalization of the kinetic term of the nilpotent chiral superfield $X$ gets modified to
\begin{equation}
\label{GEN2}
\int d^4 \theta \left( \frac{{\cal F} + \overline {\cal F}}{2} \right) X \overline X .
\end{equation} 

\subsection{Mixed F- and D-term breaking} 

As a final example, we consider a case that involves a mixing between F- and D-term breaking contributions.
A simple model of this type is described by the following Lagrangian 
\begin{equation}
\label{ggL}
\begin{aligned}
\mathcal{L} = &  
\frac{1}{4} \left(\int d^2\theta \, \left\{1 + \frac{\Phi}{M} \right\} \,  W^\alpha W_\alpha +c.c.\right) 
+ \xi \int d^4 \theta\, V 
\\ 
& + \int d^4 \theta \Phi \overline \Phi  + \left(\int d^2 \theta\, \left(f \Phi+\frac{m}{2}\Phi^2\right) + c.c. \right) \,,
\end{aligned}
\end{equation}
where the parameters $f$, $\xi$, $m$ and $M$ are chosen to be real for simplicity.
At the component level, the F-term and D-term contributions to the scalar potential are respectively
\begin{equation}
{\cal V}_\text{F} = f^2 +2\,mfa +m^2(a^2+b^2)\ , \qquad {\cal V}_\text{D} = \frac{ \xi^2}{8\left(1+\frac{a}{M}\right) } \, , 
\end{equation}
where we decomposed the scalar component of $\Phi$ into its real and imaginary parts  $\phi = a+ i b$. 
The purpose of this example is to get a low energy effective theory in which the goldstino is interacting with the vector, like in the case of pure D-term breaking. 
The difference with respect to that example, however, is that this time both the gaugino and the fermion in $\Phi$ will get a mass. 
The goldstino will be then the massless combination of this two fermions, while the massive combination is going to decouple in the low energy limit. 

We will now differentiate between two different possible regimes of supersymmetry breaking, one where the F-term is dominating and one where the D-term is dominating.
While some details of the two regimes change, we will see that we can rather easily cover both examples by the same parametrization of the linear multiplets in terms of the same constrained superfields.

The regime where the F-term source of supersymmetry breaking is dominating is obtained when 
\begin{equation}
M  \gg \sqrt{f} \gg \sqrt \xi \gg E,
\end{equation}
where $E$ is expressing the energy range of validity of the effective theory, while the linear theory (\ref{ggL}) is valid for $ M \gg E > \sqrt{f}$.
For simplicity, given the aforementioned hierarchy and since we are interested in the qualitative behaviour of the model in the low energy, we will tune the parameters such that
\begin{equation}
{\xi}=4\sqrt{M^3 m}, \qquad f=M^2.
\end{equation}
With this particular choice the potential has a minimum at
\begin{equation}
\langle a\rangle = 0, \qquad \langle b \rangle =0.
\end{equation}
In this minimum the mass spectrum is made up of one massless vector, two real scalars of mass $m_a^2=4m^2 + 4mM$ and $m_b^2 = 2m^2$ respectively, one goldstino and one massive fermion of mass $m_f^2 = \frac{(M-2m)^2}{16}$.
We therefore expect that at low energies only the vector field and the goldstino survive.
The goldstino and the massive fermion are given respectively by the following combinations of $\psi^\Phi$, the fermion in $\Phi$, and $\lambda$, the gaugino
\begin{equation}
\label{eigenvectors}
\begin{aligned}
&-\frac{i}{\sqrt 2}\sqrt\frac{M}{m}\psi^\Phi +\lambda \quad \sim \quad \text{goldstino},\\[3mm]
&-i\sqrt{2}\sqrt{\frac{m}{M}}\psi^\Phi + \lambda \quad \sim \quad \text{massive fermion}.
\end{aligned}
\end{equation}
We see that, since $M \gg m$, the goldstino mostly resides in $\Phi$.

On the other hand, we will have D-term dominated supersymmetry breaking when
\begin{equation}
M \gg \sqrt{\xi} \gg \sqrt f \gg E.
\end{equation}
Once again, for simplicity, we can cover this case by setting
\begin{equation}
{\xi}=4\sqrt{m^3 M}, \qquad f=m^2.
\end{equation}
With this particular choice the potential has a minimum at
\begin{equation}
\langle a\rangle = 0, \qquad \langle b \rangle =0.
\end{equation}
In this minimum the mass spectrum is made up of one massless vector, two real scalars of mass $m_a^2=\frac{2m^2(2m+M)}{M}$ and $m_b^2 = 2m^2$, one goldstino and one massive fermion of mass $m_f^2 = \frac{m^2(m-2M)^2}{16 M^2}$.
Also in this case we expect that in the low energy limit only the vector field and the goldstino will survive. 
The goldstino and the massive fermion are given respectively by the following combinations of $\psi^\Phi$, the fermion in $\Phi$, and $\lambda$, the gaugino 
\begin{equation}
\label{eigenvectors2}
\begin{aligned}
&-\frac{i}{\sqrt 2}\sqrt\frac{m}{M}\psi^\Phi +\lambda \quad \sim \quad \text{goldstino},\\
&-i\sqrt{2}\sqrt{\frac{M}{m}}\psi^\Phi + \lambda \quad \sim \quad \text{massive fermion}.
\end{aligned}
\end{equation}
In this case the goldstino mostly resides in $V$, because $M \gg m$. 

In principle various different parametrizations of the unconstrained superfields in terms of the constrained ones can be adopted. 
To obtain the desired decoupling, however, it is convenient to introduce the following parametrization 
\begin{equation}
\label{ggR}
\begin{aligned}
V & = \hat V + \sqrt 2 \frac{ Y \overline X }{\overline D^2 \overline X} + \sqrt 2 \frac{ \overline Y X }{D^2 X} \,  ,
\\
\Phi & = X + S \, ,
\end{aligned}
\end{equation} 
where $Y$ and $\hat V$ are respectively a chiral and a vector superfields satisfying 
\begin{equation}
X Y = 0 \, , \qquad X \overline X D^2 Y  = 0  
\, , \qquad  X \hat W_\alpha = 0  \, ,  
\end{equation}
while the chiral superfields $X$ and $S$ satisfy  \eqref{XH1}. 
In particular the only independent component of $Y$ is the fermion $\psi$, which is going to be aligned with the massive fermion for these models.
By inserting the parametrization \eqref{ggR} in the Lagrangian \eqref{ggL} the model can be written entirely in terms of constrained superfields as
\begin{equation}
\label{ggL2}
\begin{aligned}
\mathcal{L} = &  
\frac{1}{4} \left(\int d^2\theta \, \left\{1 + \frac{X + S}{M} \right\} \,  
W^2 (\hat V, Y, X) +c.c.\right) 
+ \xi \int d^4 \theta\, \hat V 
- \frac{\sqrt 2 \xi}{4} \left( \int d^2 \theta \, Y + c.c. \right) 
\\[3mm]
& + \int d^4 \theta \left(X\overline X + S \overline S\right) 
+  \left(\int d^2 \theta \, \left(f (X + S)+\frac{m}{2}(XS+S^2)\right)+c.c.\right) \, , 
\end{aligned}
\end{equation} 
where the vector superfield $V$ in $W_\alpha$ is parametrized as in \eqref{ggR}.

In the zero momentum limit, this Lagrangian reduces to
\begin{equation}
\begin{aligned}
\mathcal{L} &=-f^2 -\frac{\xi^2}{8} + |F+m\overline s+f|^2+\frac12\left(\text D +\frac{\xi}{2}\right)^2-m^2 s \overline s 
\\[2mm]
& +\frac{1}{4M}\left[\text D^2- 4 m M\, f\right](s+\overline s)
+ \frac{1}{16 M}\left(F\psi^2 + \overline F\overline \psi^2\right).
\end{aligned}
\end{equation}
Both in the D-term dominated scenario and in the F-term dominated one, we can see that the minimization of the action is obtained by setting 
\begin{equation}
F = - f, \qquad \text D = -\frac{\xi}{2} = -2\sqrt{m M \, f}, \qquad s=0,
\end{equation}
because of the large mass for $s$, and this also introduces an effective large mass for $\psi$, which is then stabilized at
\begin{equation}
\psi=0.
\end{equation}
In the effective theory we can therefore set $S = Y = 0$, which are directly implied by $s = \psi =0$.

While the decoupling of the scalar $s$ is somehow straightforward, the decoupling of the massive fermion needs to be commented a bit more. 
Because of the constraint $X\hat W_\alpha = 0$ we imposed, the fermion $\hat \lambda_\alpha$ will be removed from the spectrum and expressed as
\begin{equation}
\hat \lambda_\alpha = i\, {\rm D}\frac{G_\alpha}{\sqrt{2}F}+\dots,
\end{equation}
where dots stand for terms with derivatives. 
The fermion $\lambda_\alpha$ in $V$, which is the gaugino of the linearly realized theory, will have then a contribution coming from $\hat \lambda$, one coming from $G$ and one coming from $\psi$, due to the fact that we used the parametrization \eqref{ggR}.
In the theory there will be a massive fermion and a goldstino which are given in (\ref{eigenvectors}), (\ref{eigenvectors2}).
However due to the Lagrangian we have, it turns out $\psi$ is aligned with the massive fermion (in the zero momentum limit) and therefore when we remove it from the spectrum, what remains is automatically aligned with the goldstino.

The resulting low-energy model is then
\begin{equation}
\mathcal{L}=\frac14 \left(\int d^2 \theta\, \hat W^2 +c.c.\right) + \xi \int d^4\theta \,\hat V+\int d^4\theta\, X\overline X+f\left(\int d^2\theta\, X + c.c.\right)
\end{equation}
and describes a massless vector interacting with a goldstino. 
This Lagrangian is similar to \eqref{LLXV}, but it has been obtained from a very different UV model.

\section{Discussion}

In this work we studied four-dimensional, $N=1$ globally supersymmetric gauged chiral models where D-term and F-term source the breaking of supersymmetry. 
We showed that the constrained superfields approach can capture all the properties of their low energy theories and, in particular, that the goldstino interactions can always be described by means of a nilpotent chiral superfield $X$. 
Our findings therefore strongly support the use of constrained superfields for the description of spontaneously broken supersymmetric theories. 

We would like now to point out the limitations of our results.
First, we have studied only chiral and vector superfields as the source of supersymmetry breaking, which  does not exhaust all the possibilities. 
It is known, for example, that supersymmetry breaking can be sourced also by linear superfields \cite{Kuzenko:2011ti,Farakos:2013zsa,Farakos:2015vba}. 
One could then investigate what are the general properties of this other class of models, in the spirit of the presentation we gave in this work. 
Second,  we have not discussed the consequences of coupling the system to supergravity. 
We expect that our approach can be extended in a consistent way, but following this direction is beyond the scope of the current work. 
Finally, if supersymmetry is never restored within a field theory, then it is not clear how generic the description in terms of constrained superfields can be. 
Work in this direction can be already found in \cite{Bergshoeff:2015jxa,Kallosh:2015nia,Vercnocke:2016fbt,Kallosh:2016aep,Aalsma:2017ulu,Garcia-Etxebarria:2015lif,Dasgupta:2016prs}. 

\section*{Acknowledgments}

\noindent 
We thank Athanasios Chatzistavrakidis, Alex Kehagias, Luca Martucci, Diederik Roest, Antoine Van Proeyen, Pantelis Tziveloglou and especially Fabio Zwirner for discussions. 
This work is supported in parts by the Padova University Project CPDA144437. 

\appendix

\section{Equations of motion for $\mathcal{A}_2$} 
\label{appA2}

In this appendix we prove that the equations of motion stemming from the Lagrangian \eqref{LAA12} gives
\begin{equation}
\mathcal{A}_2 =0.
\end{equation}
Since we are going to take the variation with respect to $\mathcal{A}_2 $, we focus only on the part of \eqref{LAA12} which depends on this quantity and we implement the constraint \eqref{constrXAZA} via a complex Lagrange multiplier $C$. 
Moreover we insert a generic non-vanishing real function $U$ in front of the term $Z \overline Z |\mathcal{A}_2|^2$, because we would like to use the result we are going to prove also in more general cases like when the canonical K\"ahler potential $X\overline X$ is multiplied by the real part of the gauge kinetic function $\mathcal{F}(\Phi)$ (see formula \eqref{GEN2}). 

The Lagrangian we start from is therefore
\begin{equation}
\mathcal{L} = \int d^4 \theta\, U \, Z\overline Z \, \mathcal{A}_2 \overline{\cal A}_2 + \int d^4\theta  \left[C (Z\mathcal{A}_2-Z\overline {\cal A}_2)+c.c.\right]
\end{equation} 
and $Z$ satisfies also the constraints \eqref{constrZ}. 
For the rest of the proof we are going to omit the subscript on $\mathcal{A}_2$. 
Taking the variations with respect to $C$ and to $\mathcal{A}$ gives the following superspace equations of motion
\begin{equation}
\begin{aligned}
&Z\mathcal{A}-Z\overline{\cal A}=0,\\
&\overline D^2\left[ Z \overline Z \mathcal{A} \,U+CZ-\overline C\overline Z\right]=0.
\end{aligned}
\end{equation}
Acting on the second equation with $\overline{Z}$ and dividing by the invertible quantity $\overline D^2 \overline Z$ gives
\begin{equation}
\overline Z \overline C = Z \overline Z \mathcal{A} U+\overline Z \overline  D^2 (ZC),
\end{equation}
where we used also the constraints \eqref{constrZ} and \eqref{constrXAZA}. 
We start now replacing iteratively this relation into its complex conjugated obtaining
\begin{equation}
\begin{aligned}
ZC &=Z \overline Z \mathcal{A} U+Z   D^2 (\overline Z\overline C)\\
&=2Z \overline Z \mathcal{A} U+\overline Z \overline  D^2 (ZC)\\
&=3Z \overline Z \mathcal{A} U+Z   D^2 (\overline Z\overline C).
\end{aligned}
\end{equation}
By comparing the first line with the last we get
\begin{equation}
Z \overline Z \mathcal{A} U = 0,
\end{equation}
which reduces to $Z \overline Z \mathcal{A}=0$, because $U$ is invertible by assumption. 
Acting with superspace derivatives we can simplify also the invertible quantities $D^2 Z$, $\overline D^2 \overline Z$ and obtain finally
\begin{equation}
\mathcal{A}=0.
\end{equation}
This result has been derived from a projection of the original equation of motion, but if it solves the full system of equations, it is going to be the most general solution. 

We note that by substituting the chiral projector $\overline D^2\rightarrow (\overline{\cal D}^2-8{\cal R})$, the proof works directly in supergravity. 

\section{From Volkov--Akulov to nilpotent $X$}

We use this second appendix to explain how to go from a theory interacting with the Volkov--Akulov fermion to a theory interacting with the nilpotent chiral superfield $X$.
Our findings show that all the couplings of the VA fermion can be described by models involving the nilpotent superfield $X$.
These theories generically contain superspace higher derivatives and therefore might not be described only by a K\"ahler potential and a superpotential.

One can describe the interactions of the VA fermion by using the spinor superfield $\Gamma_\alpha$ \cite{Samuel:1982uh}, which has the properties
\begin{equation}
\begin{aligned}
D_\alpha \Gamma_\beta  &= \epsilon_{\beta \alpha} \, ,
\\
\overline D^{\dot \beta} \Gamma^\alpha &=
2 i \, \overline \sigma^{m \, \dot \beta \beta} \, \Gamma_\beta  \,
\partial_m \Gamma^\alpha \, ,
\end{aligned}
\end{equation}
where we absorbed the scale $\kappa^{-1/2}$ appearing in \cite{Samuel:1982uh} into $\Gamma_\alpha$.
A generic Lagrangian containing $\Gamma_\alpha$ will have the superspace form
\begin{equation}
\label{apLD1}
{\cal L} = - \frac{1}{16 \, \kappa^2} \int d^4  \theta \, \Gamma^2
\overline \Gamma^2
+  {\cal L}_m \left[  \Gamma, \overline \Gamma,  \text{matter
superfields} \right] \, ,
\end{equation}
where ${\cal L}_m$ is a superspace Lagrangian describing the interactions of the VA fermion with matter.
Note that, even if the Lagrangian is not in the form \eqref{apLD1} one can add and subtract the first term to bring the model in this form.
By using the superfield $Z$ defined from \eqref{constrZ}, we have that
\cite{Cribiori:2016hdz}
\begin{equation}
\Gamma_\alpha = -2 \, \frac{D_\alpha Z}{D^2 Z} .
\end{equation}
Therefore the Lagrangian \eqref{apLD1} can be written as
\begin{equation}
\label{aptotot}
{\cal L} =  {\cal L}_Z  +  {\cal L}_m \left[
\left( -2 D_\alpha Z / D^2 Z \right),
\left( -2 \overline D_{\dot \alpha}  \overline Z / \overline D^2
\overline Z \right),
\text{matter superfields} \right] \, ,
\end{equation}
where
\begin{equation}
\label{apfreeZZ}
{\cal L}_Z = - \kappa^{-2} \int d^4 \theta \, Z \overline Z .
\end{equation}
As we will see in the following considerations, the matter Lagrangian in \eqref{aptotot} is irrelevant and we can safely ignore it when proving the equivalence, but we will restore it in the final result.
Therefore we focus only on \eqref{apfreeZZ} and we will prove it is equivalent to the free Lagrangian for the $X$ superfield \eqref{VA}.

To prove the equivalence between \eqref{apfreeZZ} and \eqref{VA} we consider the following superspace Lagrangian
\begin{equation}
\label{apLD}
\kappa^2 {\cal L}_{0} = \int d^4 \theta \, Z \overline Z \left( D^2 X \,
\overline D^2 \overline X -  D^2 X - \overline D^2 \overline X \right)
+ \left\{ \int d^4 \theta \, J \left( X - \frac{Z}{D^2 Z} D^2 X
\right) + c.c. \right\} \, .
\end{equation}
Here the superfield $J$ is complex and completely unconstrained, while $X$ is an unconstrained chiral superfield.
Notice that the Lagrangian part containing matter in \eqref{aptotot} does not contain $X$, neither $J$, therefore any variation with respect to these superfields is not affected by that part of the theory.
We will show that by integrating out $J$ from \eqref{apLD} one finds the Lagrangian \eqref{VA} for a nilpotent $X$ with $X^2=0$, whereas integrating out $X$ gives the Lagrangian \eqref{apfreeZZ} for $Z$.

Let us first show that \eqref{apLD} is equivalent to \eqref{apfreeZZ}.
To this end we vary the unconstrained but chiral superfield $X$ and we find
\begin{equation}
\label{apEOM}
\overline D^2 \left[ D^2 \left( Z \overline Z \, \overline D^2
\overline X - Z \overline Z \right)
+ J
- D^2 \left( J \frac{Z}{D^2 Z} \right) \right] = 0 \, .
\end{equation}
Once we  multiply \eqref{apEOM}  with $Z \overline Z$ we find
\begin{equation}
\label{apeom1}
Z \overline Z \, D^2 X =  Z \overline Z \, ,
\end{equation}
which combined with \eqref{apEOM} gives
\begin{equation}
\label{apeom2}
\overline D^2 \left[ J
- D^2 \left( J \frac{Z}{D^2 Z} \right) \right] = 0 .
\end{equation}
By inserting \eqref{apeom1} and \eqref{apeom2} into \eqref{apLD} we get
\begin{equation}
\label{apBB}
{\cal L}_0 =  - \kappa^{-2} \int d^4 \theta \, Z \overline Z
= - \frac{1}{16 \, \kappa^{2}} \int
d^4 \theta \, \Gamma^2 \overline \Gamma^2 \, .
\end{equation}

We can also show that \eqref{apLD} is equivalent to \eqref{VA} for a nilpotent $X$.
Once we vary the unconstrained $J$ we find
\begin{equation}
\label{apX33}
X = Z\frac{D^2 X }{D^2 Z} .
\end{equation}
From \eqref{apX33} we see that
\begin{equation}
\label{X2sol}
X^2 = 0 \qquad \Longrightarrow \qquad X=-\frac{D^\alpha X D_\alpha X}{D^2 X}\,,
\end{equation}
but also that
\begin{equation}
\label{apDXDPhi}
\frac{D_\alpha X}{D^2 X} = \frac{D_\alpha  Z}{D^2 Z} \, .
\end{equation}
Finally from \eqref{X2sol} and \eqref{apDXDPhi} we find
\begin{equation}
\label{apXP}
Z \overline Z \, D^2 X
= \frac{D^\alpha Z D_\alpha Z}{(D^2 Z)^2} \frac{\overline D_{\dot\alpha} \overline Z \, \overline
D^{\dot\alpha} \overline Z}{(\overline D^2 \overline Z)^2} \, D^2 X
=  \frac{D^\alpha X D_\alpha X}{(D^2 X)^2} \frac{\overline D_{\dot\alpha} \overline X \,
\overline D^{\dot\alpha} \overline X}{(\overline D^2 \overline X)^2} \, D^2 X
= \frac{X \overline X}{\overline D^2 \overline X} .
\end{equation}
Using \eqref{apXP} the Lagrangian \eqref{apLD}  becomes
\begin{equation}
\label{apB}
{\cal L}_0 = \kappa^{-2} \int d^4 \theta \, X \overline X
+ \left( \frac{1}{4 \, \kappa^2} \int d^2 \theta \, X + c.c. \right) \, .
\end{equation}
Notice that here $X$ has mass dimensions $[X]=-1$, but we can restore the correct dimension by rescaling it with $\kappa$.

Once we perform the above procedure taking into account the full Lagrangian \eqref{aptotot} and after the rescaling of $X$ with $\kappa$ we will have
\begin{equation}
\label{apXLXL}
\begin{aligned}
{\cal L} = &  \int d^4 \theta \, X \overline X +   \left( \frac{1}{4
\kappa} \int
d^2 \theta \, X + c.c. \right)
\\
& + {\cal L}_m \left[
\left( -2 D_\alpha X / D^2 X \right),
\left( -2 \overline D_{\dot \alpha}  \overline X / \overline D^2
\overline X \right),
\text{matter superfields} \right] \, .
\end{aligned}
\end{equation}
The equivalence between \eqref{apBB} and \eqref{apB} was proved in superspace earlier in \cite{Cribiori:2016hdz}, but here our proof goes through also when we include the possible matter couplings and we expect it to hold also in supergravity.

\end{document}